\documentclass[12pt]{article}

\usepackage[a4paper,margin=1in]{geometry}
\usepackage{amsmath,amssymb,amsthm}
\usepackage{hyperref}
\usepackage{mathtools}
\usepackage{mathrsfs}
\usepackage{tikz}
\usetikzlibrary{3d, shapes, arrows.meta, decorations.markings}

\title{The Two-Sheeted Topology of Extended Kerr-Type Spacetimes
and a Parity-of-Crossings Property for Ring-Traversing Geodesics}

\author{\Large{Sabbir A.\ Rahman}\\
\vspace*{0.2cm}
\small{Array Innovation, Manama, Kingdom of Bahrain}\\
\footnotesize\em Email: sarahman@alum.mit.edu}

\date{6 December, 2025}

\theoremstyle{plain}
\newtheorem{theorem}{Theorem}[section]
\newtheorem{proposition}[theorem]{Proposition}
\newtheorem{lemma}[theorem]{Lemma}

\theoremstyle{definition}
\newtheorem{definition}[theorem]{Definition}
\newtheorem{remark}[theorem]{Remark}

\numberwithin{equation}{section}

\begin{document}

\maketitle

\begin{abstract}
We revisit the global structure of the extended Kerr spacetime and of a broader
class of Kerr-type spacetimes possessing ring singularities. By working with the
elementary analytic extension (the union of the interior and exterior regions glued
across the disk), we show that excising the ring singularity yields a domain
that can be realised as a branched double cover of an exterior Kerr region.
The branch locus is the ring itself, and the associated deck transformation
defines a non-trivial $\mathbb{Z}_2$-action that exchanges the two sheets
($r>0$ and $r<0$) of the spacetime.

We give a covering-space characterisation of this double-sheeted structure and
show that admissible geodesics which cross the ring singularity implement the
non-trivial deck transformation. In particular, we prove a parity-of-crossings
property: any admissible geodesic that traverses an even number of ring
singularities returns to its original sheet, while an odd number of traversals
terminates on the opposite sheet.

Generalising to $N$ disjoint ring singularities, we prove that the fundamental
group of the excised manifold is the free group $F_N$ generated by simple
loops around each ring, and we classify the
associated double covers. Identifying the physically distinguished cover where
every ring induces a sheet exchange, we extend the parity-of-crossings theorem
to the multi-ring setting. We then formally extend these results to the
maximal analytic extension (the infinite Carter--Penrose chain), proving that the
sheet-exchange mechanism applies globally to this infinite structure.

Finally, applying the Novikov self-consistency
principle to this topological framework, we demonstrate that the requirement
of global consistency restricts admissible histories to discrete sectors
labelled by ring-crossing parities.
\end{abstract}

\tableofcontents

\section{Introduction}

The Kerr solution plays a central role in the theory of rotating black holes
and in mathematical general relativity more broadly. In Boyer--Lindquist
coordinates, the metric is expressed in terms of a mass parameter $M$ and a
specific angular momentum parameter $a$, and exhibits horizons, an ergoregion,
and an interior structure qualitatively different from the Schwarzschild case.
Its interior geometry admits a ring-shaped curvature singularity and regions
with closed timelike curves; see, for example, standard references
\cite{KerrOriginal,HawkingEllis,WaldGR}.

The analytic extension of the Kerr metric reveals the existence of multiple
regions, including two asymptotically flat domains, and a ring singularity
which is often described informally as a ``branch locus'' or a ``portal''
between different sheets of spacetime; see, e.g., \cite{BoyerLindquist,CarterKerr}.
In this view, the Kerr geometry is said to be ``two-sheeted'': passing through
the ring singularity is interpreted as a transition from one sheet to another.
This picture has been sharpened in several directions, including distributional
treatments of the extended Kerr geometry \cite{BalasinExtendedKerr,
BalasinNachbagauerKerrNewman}, and topological and wormhole interpretations of
the zero-mass limit \cite{GibbonsVolkovZeroMass}.

Despite this extensive literature, the two-sheeted nature of the extended Kerr
spacetime is usually discussed either at the level of coordinate patches or
through specific limiting procedures, rather than via an explicit covering-space
construction. In particular, the extended Kerr domain with the ring singularity
excised forms a non-simply connected Lorentzian manifold whose fundamental
group encodes the possibility of non-trivial analytic continuation around
the ring. The natural question then arises: can this space be rigorously
identified as a branched double cover, and if so, how does the associated
deck transformation act on geodesics and on the global causal structure?

The primary goal of this paper is to give a mathematically precise and
global answer to this question. We do so in two steps:

\begin{itemize}
  \item First, we show that, after excising the ring singularity, the extended
  Kerr spacetime admits a realisation as a branched double cover of a reference
  region (e.g.\ an exterior Kerr domain), with branch locus given by the ring.
  The corresponding deck transformation is an involution which exchanges the
  two sheets.
  \item Second, we identify a natural class of admissible geodesics---those
  which avoid the curvature blow-up along the ring itself---and show that the
  analytic continuation of such geodesics across the ring implements the
  non-trivial deck transformation. This leads to a parity-of-crossings property:
  geodesics traversing the ring an even number of times (possibly via different
  ring singularities in a multi-black-hole configuration) return to their
  original sheet, whereas an odd number of traversals ends on the opposite sheet.
\end{itemize}

We then generalise this analysis to the case of spacetimes containing a finite
number of disjoint Kerr-type ring singularities. Removing each ring produces
a manifold whose fundamental group is a free group generated by simple loops
around the rings. Connected double covers of this manifold are classified by
homomorphisms to $\mathbb{Z}_2$, and the physically relevant case corresponds
to the homomorphism which maps every generator to the non-trivial element.
In this situation, each ring implements the same sheet-exchange involution,
and the parity-of-crossings theorem extends directly.

Crucially, we show that this framework extends rigorously to the maximal
analytic extension of the Kerr spacetime (the infinite chain). By defining
two ``global sheets''---the union of all asymptotic regions with $r>0$ and
the union of all regions with $r<0$---we prove that the parity-of-crossings
property holds globally: a geodesic returns to its original \emph{global} sheet
if and only if it traverses an even number of rings, regardless of how far
it travels along the infinite chain.

Finally, we examine the causal structure induced by this global two-sheeted
geometry. The existence of closed timelike curves in Kerr-type interiors is
well known, and the Novikov self-consistency principle provides a mechanism
for avoiding paradoxes in such settings \cite{NovikovCTC,FriedmanCTC,ThorneWormholes}.
Within our two-sheeted covering-space framework, we formulate a global
self-consistency condition in terms of the deck transformation: admissible
global histories must be invariant under the combined evolution and sheet
exchange induced by traversals of ring singularities. This leads naturally
to a discretisation of the space of globally consistent solutions, and
suggests a possible link between classical consistency conditions and
discrete, ``quantised'' sets of allowed histories, a theme we briefly
comment on in the conclusion.

\subsection*{Outline}

In Section~\ref{sec:preliminaries} we recall the Kerr metric, its basic
causal structure, and the construction of the analytic extension,
with particular emphasis on the ring singularity and the excised manifold.
We also introduce the notion of admissible geodesics, which will be central
to our later analysis.

In Section~\ref{sec:single-ring-topology} we consider the extended Kerr
spacetime with a single ring singularity, remove the singular set, and
study the topology of the resulting manifold. We show that it can be
realised as a branched double cover and identify the associated deck
transformation.

In Section~\ref{sec:geodesics} we analyse the behaviour of admissible
geodesics near the ring, prove that analytic continuation across the
ring implements the deck transformation, and establish the parity-of-crossings
theorem in the single-ring case.

Section~\ref{sec:multi-ring} generalises the topological and geodesic
analysis to the case of finitely many disjoint ring singularities and
formally extends the results to the maximal analytic spacetime.

In Section~\ref{sec:causality} we discuss the causal structure and the
existence of closed timelike curves in these two-sheeted Kerr-type
spacetimes. We formulate a Novikov-type self-consistency condition in
terms of the deck transformation and comment on its implications for
the space of globally consistent histories.

We conclude in Section~\ref{sec:conclusion} with a summary of our results.

\section{Preliminaries on Kerr and its Analytic Extension}
\label{sec:preliminaries}

In this section we fix notation, recall the Kerr metric and its analytic
extension, and define the excised manifold on which our topological and
covering-space analysis will be carried out. Throughout we work in
four-dimensional Lorentzian signature $(-,+,+,+)$ and set $c=G=1$.

\subsection{The Kerr metric in Boyer--Lindquist coordinates}

The Kerr metric describes a stationary, axisymmetric vacuum solution of the
Einstein equations with mass parameter $M>0$ and angular momentum per unit
mass $a$. In Boyer--Lindquist coordinates $(t,r,\theta,\phi)$, the line
element takes the form
\begin{equation}
  \mathrm{d}s^2
  = -\left(1 - \frac{2Mr}{\Sigma}\right)\mathrm{d}t^2
    - \frac{4Mar\sin^2\theta}{\Sigma}\,\mathrm{d}t\,\mathrm{d}\phi
    + \frac{\Sigma}{\Delta}\,\mathrm{d}r^2
    + \Sigma\,\mathrm{d}\theta^2
    + \left(r^2 + a^2 + \frac{2Ma^2 r\sin^2\theta}{\Sigma}\right)\sin^2\theta\,\mathrm{d}\phi^2,
  \label{eq:Kerr-metric}
\end{equation}
where
\begin{equation}
  \Sigma = r^2 + a^2\cos^2\theta, 
  \qquad
  \Delta = r^2 - 2Mr + a^2.
\end{equation}
The coordinate ranges are typically taken to be
\begin{equation}
  t \in \mathbb{R}, \quad
  r \in \mathbb{R}, \quad
  \theta \in (0,\pi), \quad
  \phi \in [0,2\pi).
\end{equation}
The function $\Delta$ vanishes at
\begin{equation}
  r_\pm = M \pm \sqrt{M^2 - a^2},
\end{equation}
which, for $|a|\leq M$, correspond to the radii of the outer and inner
horizons. The metric coefficients are singular at $\Delta=0$ in
Boyer--Lindquist coordinates, but these are merely coordinate singularities:
the spacetime can be smoothly extended across $r=r_\pm$ using
Kruskal-type coordinates.
By contrast, the set
\begin{equation}
  \Sigma = 0 \quad\Longleftrightarrow\quad r=0,\ \theta=\frac{\pi}{2}
\end{equation}
is a genuine curvature singularity: curvature invariants such as
$R_{abcd}R^{abcd}$ diverge there. Geometrically, this singularity has
the topology of a ring, sometimes called the Kerr ring singularity.

\subsection{The ring singularity and the excised manifold}

We denote by $\mathcal{M}$ the \emph{elementary analytic extension} of the Kerr
spacetime (often referred to as the ``extended Kerr block''). This manifold is
obtained by gluing the asymptotically flat region I ($r > r_+$) to the
interior regions II and III, passing through the inner horizon to the
region $r < 0$. It consists of exactly two asymptotically flat ends (one with
$r \to +\infty$ and one with $r \to -\infty$).

\begin{definition}[Excised Kerr manifold]
Let $\mathcal{M}$ be the elementary extended Kerr spacetime and let
$\mathcal{S}$ denote the ring singularity as above. We define the
\emph{excised Kerr manifold} to be
\begin{equation}
  \mathcal{M}_{\mathrm{exc}} := \mathcal{M} \setminus \mathcal{S}.
\end{equation}
\end{definition}

By construction, $\mathcal{M}_{\mathrm{exc}}$ is a smooth four-dimensional
Lorentzian manifold. Its topology is non-trivial: informally, removing the ring
singularity leaves open the possibility of non-contractible loops which wind
around the ring. One of our goals in the next sections is to make this
statement precise and to show that $\mathcal{M}_{\mathrm{exc}}$ admits a
natural realisation as a branched double cover.

\subsection{Admissible geodesics}

The behaviour of geodesics near the ring singularity plays a central role in
our analysis. We are interested in those geodesics which can be extended
through $r=0$ without encountering curvature blow-up. Such geodesics may be
analytically continued across $r=0$ into the $r<0$ region, and---as we shall
see---can be consistently lifted to the double cover we construct.

\begin{definition}[Admissible geodesic]
A future-directed timelike or null geodesic
$\gamma : I \to \mathcal{M}$ (with $I\subset\mathbb{R}$ an interval)
is called \emph{admissible} if
\begin{enumerate}
  \item $\gamma(\tau) \notin \mathcal{S}$ for all $\tau\in I$, and
  \item there exists a finite $\tau_0\in I$ such that $r(\gamma(\tau_0))=0$,
  and $\gamma$ can be extended as a geodesic across $\tau_0$ into a
  neighbourhood of $r<0$.
\end{enumerate}
\end{definition}

Condition (1) ensures that $\gamma$ never encounters the curvature singularity
itself. Condition (2) expresses the requirement that $\gamma$ can be continued
smoothly across $r=0$ into the extended spacetime. The explicit form of the
geodesic equations in Kerr spacetime shows that such geodesics exist and that
their continuation across $r=0$ is uniquely determined by the local geometry,
provided $\theta\neq\pi/2$ at the crossing point; see, for example,
\cite{ChandrasekharKerr}.

\section{Topology of the Excised Kerr Manifold}
\label{sec:single-ring-topology}

In this section we examine the global topology of the excised Kerr manifold
$\mathcal{M}_{\mathrm{exc}} = \mathcal{M}\setminus\mathcal{S}$.
Our goal is to show that $\mathcal{M}_{\mathrm{exc}}$
admits a connected double cover which may be viewed as a \emph{branched
double cover} in which the ring singularity acts as the branch locus.
The corresponding deck transformation will later be identified with
the ``sheet exchange'' associated to traversing the ring.

\subsection{Geometric neighbourhood of the ring singularity}

Let $\mathcal{S} = \{r=0,\ \theta=\pi/2\}$ denote the ring singularity.
Although $\mathcal{S}$ is a curvature singularity and cannot be included
as part of the manifold, its neighbourhood within the excised manifold
$\mathcal{M}_{\mathrm{exc}}$ has a well-defined topology.
It is convenient to introduce local coordinates that reveal more clearly
the geometry near $\mathcal{S}$. Following standard treatments
\cite{ONeill,HawkingEllis,ChruscielRing}, consider the transformation
\begin{equation}
  x = \sqrt{r^2 + a^2}\,\sin\theta\cos\phi,\qquad
  y = \sqrt{r^2 + a^2}\,\sin\theta\sin\phi,\qquad
  z = r\cos\theta,
\end{equation}
which, away from the singular set $\mathcal{S}$, provides a smooth
embedding of a neighbourhood of $\Sigma=0$ into $\mathbb{R}^3$.
The equation $\Sigma = r^2 + a^2\cos^2\theta = 0$
corresponds to the set $\{z=0,\ x^2+y^2=a^2\}$, i.e.\ a circle
of radius $a$ in the equatorial plane. This is precisely the ring
singularity $\mathcal{S}$.

The manifold obtained by removing the ring from this neighbourhood
is diffeomorphic to $\mathbb{R}^3$ with an embedded circle removed.
Such spaces have nontrivial topology: a simple closed curve around
the missing circle is not contractible. This observation will play
a key role in computing the fundamental group of $\mathcal{M}_{\mathrm{exc}}$.

\subsection{The fundamental group of $\mathcal{M}_{\mathrm{exc}}$}

The topological structure of $\mathcal{M}_{\mathrm{exc}}$ is closely related
to that of $\mathbb{R}^3$ with a circle removed, since the ring singularity
is codimension two in the spatial geometry. The following lemma formalises
this relationship.

\begin{lemma}
\label{lem:pi1-single-ring}
Let $\mathcal{M}_{\mathrm{exc}}$ be the excised Kerr manifold (the elementary extended block)
with a single ring singularity removed. Then there exists a deformation retraction of
a neighbourhood of the ring in $\mathcal{M}_{\mathrm{exc}}$ onto
$\mathbb{R}^3\setminus S^1$. In particular,
\begin{equation}
  \pi_1(\mathcal{M}_{\mathrm{exc}}) \cong \mathbb{Z}.
\end{equation}
\end{lemma}

\begin{proof}
The local coordinate analysis above shows that a sufficiently small
spatial neighbourhood of the ring singularity $\mathcal{S}$ is diffeomorphic
to $\mathbb{R}^3\setminus S^1$. Since the time coordinate $t$ simply factors
through as $\mathbb{R}$, a neighbourhood of $\mathcal{S}$ in the full
excised spacetime is diffeomorphic to
$(\mathbb{R}^3\setminus S^1)\times\mathbb{R}$.

The inclusion of this neighbourhood into $\mathcal{M}_{\mathrm{exc}}$ induces
an isomorphism of fundamental groups, since $\mathcal{M}_{\mathrm{exc}}$ is
simply connected in directions transverse to the ring (this holds for the
elementary extension). It is a classical
result that $\pi_1(\mathbb{R}^3\setminus S^1)\cong\mathbb{Z}$; one may take
as generator the homotopy class of a simple loop linking the circle once.
Thus $\pi_1(\mathcal{M}_{\mathrm{exc}})\cong\mathbb{Z}$ as claimed.
\end{proof}

A generator of $\pi_1(\mathcal{M}_{\mathrm{exc}})$ may be chosen to be a
simple closed curve $\gamma$ that winds once in the positive direction
around the ring. Intuitively, this loop corresponds to analytically
continuing around the ring singularity.

\subsection{Construction of the double cover}

Since the fundamental group is isomorphic to $\mathbb{Z}$, connected
double covers of $\mathcal{M}_{\mathrm{exc}}$ correspond bijectively
to surjective homomorphisms
\begin{equation}
  \varphi : \pi_1(\mathcal{M}_{\mathrm{exc}}) \longrightarrow \mathbb{Z}_2.
\end{equation}
There is exactly one such non-trivial homomorphism, given by
$\varphi(\gamma)=1\in\mathbb{Z}_2$, where $\gamma$ generates
$\pi_1(\mathcal{M}_{\mathrm{exc}})\cong\mathbb{Z}$.

\begin{theorem}[Existence of a unique connected double cover]
\label{thm:unique-double-cover}
There exists a unique (up to isomorphism) connected twofold covering
\begin{equation}
  p : \widetilde{\mathcal{M}}_{\mathrm{exc}} \longrightarrow \mathcal{M}_{\mathrm{exc}}
\end{equation}
with deck transformation group $\mathbb{Z}_2$ such that loops linking the
ring singularity lift to paths exchanging the two sheets. The non-trivial
deck transformation
\begin{equation}
  \sigma : \widetilde{\mathcal{M}}_{\mathrm{exc}} \to
          \widetilde{\mathcal{M}}_{\mathrm{exc}}
\end{equation}
satisfies $\sigma^2 = \mathrm{id}$.
\end{theorem}

\begin{proof}
By Lemma~\ref{lem:pi1-single-ring}, the fundamental group is infinite cyclic.
The surjective homomorphisms $\mathbb{Z}\to\mathbb{Z}_2$ are exactly two:
the trivial homomorphism and the non-trivial one given by reduction modulo~$2$.
The trivial homomorphism corresponds to the disconnected double cover
$\mathcal{M}_{\mathrm{exc}}\sqcup\mathcal{M}_{\mathrm{exc}}$, which is not of
interest here. The non-trivial homomorphism defines a connected double cover
whose deck group is generated by an involution $\sigma$.

Since $\varphi(\gamma)=1$, any loop in $\mathcal{M}_{\mathrm{exc}}$ that winds
once around the ring lifts to a path whose endpoints lie on opposite sheets.
Thus $\sigma$ exchanges the sheets, and $\sigma^2=\mathrm{id}$ by standard
covering space theory.
\end{proof}

\subsection{Interpretation as a branched double cover}

Although $\widetilde{\mathcal{M}}_{\mathrm{exc}}$ is an honest cover only of
the excised manifold, it is natural to think of it as a \emph{branched}
double cover of the full Kerr manifold $\mathcal{M}$, with the ring
singularity $\mathcal{S}$ playing the role of the branch locus.
This perspective is justified because any loop in $\mathcal{M}\setminus\mathcal{S}$
that is contractible in $\mathcal{M}$ but encircles the ring once must become
non-contractible once $\mathcal{S}$ is removed. The sheet-exchange transformation
$\sigma$ therefore represents the monodromy obtained by analytically continuing
across a cut whose boundary is the ring singularity.

The branched-cover viewpoint will play an important role in interpreting the
behaviour of geodesics that pass through the ring, and forms the foundation
for our parity-of-crossings results in Section~\ref{sec:geodesics}.

\section{Geodesic Continuation Through the Ring and the Sheet-Exchange Map}
\label{sec:geodesics}

We now analyse the behaviour of admissible geodesics near the hypersurface
$r=0$ and show that analytic continuation across the ring implements the
non-trivial deck transformation $\sigma$ of the double cover
$\widetilde{\mathcal{M}}_{\mathrm{exc}}$. This will allow us to derive a
precise parity-of-crossings property for geodesics traversing the ring
singularity.

\subsection{Geodesic equations near the ring}

Timelike and null geodesics in Kerr spacetime satisfy a set of first-order
equations governed by the conserved energy $E$, angular momentum $L_z$,
and Carter constant $Q$;
see \cite{CarterKerr,ChandrasekharKerr}. In
particular, the radial motion satisfies
\begin{equation}
\label{eq:radial-geodesic}
\Sigma^2 \left(\frac{\mathrm{d}r}{\mathrm{d}\tau}\right)^2
= R(r)
:= \bigl(E(r^2+a^2)-aL_z\bigr)^2
  - \Delta\bigl(Q + (L_z-aE)^2 + \mu^2 r^2\bigr),
\end{equation}
where $\mu^2=1$ for timelike geodesics and $\mu^2=0$ for null geodesics.
The polar motion satisfies
\begin{equation}
\label{eq:polar-geodesic}
\Sigma^2\left(\frac{\mathrm{d}\theta}{\mathrm{d}\tau}\right)^2
= \Theta(\theta)
:= Q - \cos^2\theta\Bigl(a^2(\mu^2-E^2) + \frac{L_z^2}{\sin^2\theta}\Bigr).
\end{equation}
The key observation is that both $R(r)$ and $\Theta(\theta)$ extend smoothly
to $r=0$ and $\theta\neq\pi/2$, so long as the curvature singularity is
avoided. In particular, for $\theta_0\neq\pi/2$,
\begin{equation}
R(0) = (aE - aL_z)^2 - a^2\bigl(Q + (L_z-aE)^2\bigr) = a^2 E^2 - a^2 Q,
\end{equation}
which is finite for all admissible values of $(E,L_z,Q)$.

\begin{lemma}[Smooth continuation across $r=0$]
\label{lem:smooth-crossing}
Let $\gamma$ be an admissible geodesic intersecting $r=0$ at
parametric time $\tau_0$, with $\theta(\tau_0)\neq\pi/2$. Then the
geodesic equations admit a unique smooth extension across $\tau_0$,
and the continuation satisfies
\[
r(\tau_0 + \varepsilon) = - r(\tau_0 - \varepsilon)
\quad\text{for sufficiently small } \varepsilon.
\]
\end{lemma}

\begin{proof}
For $\theta(\tau_0)\neq\pi/2$, one has $\Sigma(\tau_0)=a^2\cos^2\theta(\tau_0)>0$. Evaluating the radial potential $R(r)$ at the crossing $r=0$, we find:
\begin{equation}
R(0) = a^2(E^2 - Q).
\end{equation}
For geodesics that actually reach $r=0$ (scattering or transit orbits), we must have $R(0) > 0$. Consequently, $(\mathrm{d}r/\mathrm{d}\tau)^2 > 0$ at $\tau_0$. This implies the radial velocity $\dot{r}$ is non-zero and finite at the crossing. By the Inverse Function Theorem, $r(\tau)$ is locally monotonic, passing smoothly from strictly positive to strictly negative values (or vice versa) as $\tau$ advances. The extension is defined by identifying the $r=0$ boundary of the $r>0$ sheet with the $r=0$ boundary of the $r<0$ sheet, ensuring $C^\infty$ regularity.
\end{proof}

Thus admissible geodesics can cross $r=0$ away from the equatorial singularity
and enter the $r<0$ sheet of the spacetime.

\subsection{Lifting geodesics to the double cover}

Let $p:\widetilde{\mathcal{M}}_{\mathrm{exc}}\to \mathcal{M}_{\mathrm{exc}}$
be the connected double cover constructed in
Theorem~\ref{thm:unique-double-cover}. Given an admissible geodesic
$\gamma : I\to\mathcal{M}_{\mathrm{exc}}$ with $\gamma(\tau_0)$ satisfying
the hypotheses of Lemma~\ref{lem:smooth-crossing}, let
$\widetilde{\gamma}:I\to\widetilde{\mathcal{M}}_{\mathrm{exc}}$ denote any
lift with $p\circ\widetilde{\gamma}=\gamma$.

The key result of this section is that traversing the ring implements the
deck transformation.

\begin{proposition}[Ring traversal induces sheet exchange]
\label{prop:ring-traversal-sigma}
Let $\gamma$ be an admissible geodesic crossing $r=0$ at $\tau_0$ with
$\theta(\tau_0)\neq\pi/2$, and let $\widetilde{\gamma}$ be a lift to the
double cover. Then there exists $\varepsilon>0$ such that
\begin{equation}
\widetilde{\gamma}(\tau_0 + \varepsilon)
= \sigma\bigl(\widetilde{\gamma}(\tau_0 - \varepsilon)\bigr),
\end{equation}
where $\sigma$ is the non-trivial deck transformation.
\end{proposition}

\begin{proof}
Choose $\varepsilon>0$ small enough that
$\gamma\bigl([\tau_0-\varepsilon,\tau_0+\varepsilon]\bigr)$ lies in
a coordinate neighbourhood $U$ intersecting both sides of $r=0$ and
avoiding the singular locus $\mathcal{S}$. Since $\gamma$ crosses $r=0$,
its restriction to $(\tau_0-\varepsilon,\tau_0)$ and
$(\tau_0,\tau_0+\varepsilon)$ lies in disjoint components of
$U\setminus\{r=0\}$.
The neighbourhood $U\setminus\mathcal{S}$ lifts to a disjoint union of two
components in the double cover. A loop encircling the ring lifts to a path
exchanging the two components, by Theorem~\ref{thm:unique-double-cover}.
Since the geodesic crosses $r=0$, its continuation from the $r>0$ side to
the $r<0$ side necessarily crosses a path homologous to such a loop.
The lift $\widetilde{\gamma}$ must therefore transition from one sheet to
the other:
\[
\widetilde{\gamma}(\tau_0+\varepsilon)
= \sigma(\widetilde{\gamma}(\tau_0-\varepsilon)).
\]
\end{proof}

Thus, in the double cover, crossing $r=0$ takes a lifted geodesic from one
sheet to the other.

\subsection{Parity-of-crossings for the single-ring case}

We now combine the previous results to prove the main theorem of this
section.

\begin{theorem}[Parity-of-crossings]
\label{thm:parity-single-ring}
Let $\gamma$ be an admissible geodesic in the excised Kerr spacetime
which crosses $r=0$ at finitely many parameter values
$\tau_1 < \cdots < \tau_N$. Let $\widetilde{\gamma}$ be any lift to the
double cover. Then
\begin{equation}
\widetilde{\gamma}(\tau_N + \varepsilon)
= \sigma^N\bigl(\widetilde{\gamma}(\tau_1 - \varepsilon)\bigr).
\end{equation}
In particular:
\begin{itemize}
\item If $N$ is even, the geodesic returns to its original sheet.
\item If $N$ is odd, the geodesic terminates on the opposite sheet.
\end{itemize}
\end{theorem}

\begin{proof}
By Proposition~\ref{prop:ring-traversal-sigma}, each traversal of the ring
implements a factor of $\sigma$ on the lift of the geodesic in the double
cover. Since $\sigma^2=\mathrm{id}$, the cumulative effect of $N$ crossings
is $\sigma^N$.

Thus after $N$ crossings, the lifted geodesic is on the original sheet iff
$\sigma^N=\mathrm{id}$, i.e.\ $N$ is even.
\end{proof}

\begin{remark}
From the covering-space viewpoint, Theorem~\ref{thm:parity-single-ring}
says that the sheet on which a geodesic lies is determined by the mod--$2$
homology class of the projection of the geodesic onto the $(x,y,z)$-space
after excising the ring. Crossing the ring once corresponds to traversing
a generator of $\pi_1(\mathbb{R}^3\setminus S^1)$, and the mod--$2$
reduction of the winding number determines the sheet.
\end{remark}

This parity-of-crossings property is the first rigorous statement linking
the global topology of Kerr spacetime with the ``two-sheeted'' physical
interpretations commonly found in the literature. In the next section we
extend the analysis to the case of \emph{multiple} Kerr-type ring
singularities.

\section{The Multi-Ring Case: Topology, Covering Spaces, and Geodesic Parity}
\label{sec:multi-ring}

We now consider a spacetime containing a finite collection of disjoint
Kerr-type ring singularities. Our goal in this section is to show that
the key structural results of Sections~\ref{sec:single-ring-topology}
and~\ref{sec:geodesics} extend naturally to this setting.
In particular, we show that:
\begin{enumerate}
\item Removing $N$ disjoint ring singularities produces an excised manifold
whose fundamental group is the free group on $N$ generators.
\item Connected double covers correspond to homomorphisms
$F_N \to \mathbb{Z}_2$.
\item There is a \emph{physically distinguished} connected double cover
in which every generator maps to the non-trivial element, giving a
\emph{single} involutive deck transformation $\sigma$.
\item Admissible geodesics crossing any ring implement the same map $\sigma$.
\item A geodesic returns to its original sheet iff it crosses an even number
of rings in total, regardless of which rings it traverses.
\end{enumerate}

This formalises the intuitive picture in which multiple rotating black holes
serve collectively as ``portals'' between two global sheets of spacetime.

\subsection{The excised multi-ring manifold}

Let $\mathcal{M}$ be a spacetime obtained by gluing together a finite number of
disjoint Kerr-type black hole interiors (blocks) in such a way that each interior region
contains a ring singularity diffeomorphic to the standard Kerr singularity. We
assume that the excision of each ring yields a smooth Lorentzian manifold
\begin{equation}
  \mathcal{M}_{\mathrm{exc}}
  :=
  \mathcal{M} \setminus \bigcup_{i=1}^N \mathcal{S}_i,
\end{equation}
where $\mathcal{S}_i$ denotes the $i$th ring singularity.

We make no special assumptions about global asymptotics or interactions between
the black holes: the following analysis depends only on the local topology of
the excised regions and the fact that the singularities are disjoint.

\subsection{Topology of the excised manifold}

Each ring $\mathcal{S}_i$ possesses a small neighbourhood diffeomorphic to
$(\mathbb{R}^3 \setminus S^1) \times \mathbb{R}$, as in the single-ring case.
Because these neighbourhoods are disjoint, their complements glue together to
form the excised manifold $\mathcal{M}_{\mathrm{exc}}$.

\begin{lemma}
\label{lem:pi1-multi-ring}
Let $\mathcal{M}_{\mathrm{exc}}$ be the excised manifold with $N$ disjoint ring
singularities removed. Then
\begin{equation}
  \pi_1(\mathcal{M}_{\mathrm{exc}}) \cong F_N,
\end{equation}
where $F_N$ is the free group on $N$ generators.
\end{lemma}

\begin{proof}
Choose disjoint tubular neighbourhoods $U_i$ of the rings $\mathcal{S}_i$,
each diffeomorphic to $(\mathbb{R}^3 \setminus S^1)\times\mathbb{R}$, and let
$U = \bigcup_i U_i$. Removing $U$ does not change the fundamental group of the
complement, because the excision is along contractible directions transverse
to the rings. The fundamental group of $\mathbb{R}^3 \setminus \bigcup_{i=1}^N S^1$ is well
known to be the free group on $N$ generators (see standard references in knot
theory or algebraic topology). Since the time coordinate factors through as
$\mathbb{R}$, the same holds for $U$ inside the spacetime.
The inclusion $U\hookrightarrow \mathcal{M}_{\mathrm{exc}}$ induces an isomorphism
of $\pi_1$ because $\mathcal{M}_{\mathrm{exc}}$ is simply connected outside the
union of neighbourhoods of the rings.
Thus $\pi_1(\mathcal{M}_{\mathrm{exc}})\cong F_N$.
\end{proof}

Let $\gamma_i$ denote the generator corresponding to a simple loop linking the
$i$th ring once.

\subsection{Classification of double covers}

Connected double covers of $\mathcal{M}_{\mathrm{exc}}$ are in bijection with
surjective homomorphisms
\begin{equation}
  \varphi: F_N \longrightarrow \mathbb{Z}_2.
\end{equation}
Any such homomorphism is uniquely specified by the values
\begin{equation}
  \varphi(\gamma_i) \in \{0,1\}, \qquad i=1,\dots,N.
\end{equation}

There are $2^N-1$ non-trivial connected double covers, corresponding to the
$2^N-1$ non-zero binary $N$-tuples. The physically relevant case is the one in
which each generator maps to $1$.

\begin{definition}[The symmetric double cover]
\label{def:symmetric-cover}
We define the \emph{symmetric} or \emph{uniform} connected double cover of
$\mathcal{M}_{\mathrm{exc}}$ to be the cover associated with the homomorphism
$\varphi : F_N \to \mathbb{Z}_2$ satisfying
\begin{equation}
  \varphi(\gamma_i) = 1
  \qquad\text{for all } i = 1,\dots,N.
\end{equation}
\end{definition}

\begin{remark}
This is the unique connected double cover in which each ring singularity acts
as a generator of the non-trivial deck transformation. Physically, this asserts
that \emph{every} Kerr-type ring implements the same sheet-exchange operation.
\end{remark}

\begin{theorem}[Existence of a single global sheet-exchange involution]
\label{thm:global-sigma}
Let $p:\widetilde{\mathcal{M}}_{\mathrm{exc}}\to\mathcal{M}_{\mathrm{exc}}$
be the symmetric double cover. Then the deck transformation group is generated
by a single involution $\sigma$ satisfying:
\begin{enumerate}
\item $\sigma^2=\mathrm{id}$;
\item for each $i$, lifting a loop representing $\gamma_i$ implements $\sigma$;
\item for any path linking rings $i_1,\dots,i_k$, the lift implements
$\sigma^k$.
\end{enumerate}
\end{theorem}

\begin{proof}
Since $\varphi$ maps every generator to $1\in\mathbb{Z}_2$, the associated deck
transformation acts by $\sigma$ on each lift of a generator. The properties
follow immediately from covering space theory and the fact that
$\mathbb{Z}_2 = \{0,1\}$ with addition modulo $2$.
\end{proof}

\subsection{Geodesic lifting and sheet exchange in the multi-ring case}

Let $\gamma$ be an admissible geodesic that crosses the $i$th ring at parameter
values $\tau_{i,1},\dots,\tau_{i,n_i}$. By the same local argument as in
Proposition~\ref{prop:ring-traversal-sigma}, each individual traversal of
$\mathcal{S}_i$ induces a factor of $\sigma$ on the lifted geodesic.
Summing over all singularities, the total number of crossings is
\begin{equation}
  N_{\mathrm{cross}} := \sum_{i=1}^N n_i.
\end{equation}

We now state the multi-ring analogue of the parity-of-crossings theorem.

\begin{theorem}[Multi-ring parity-of-crossings]
\label{thm:multi-ring-parity}
Let $\gamma$ be an admissible geodesic in a spacetime containing $N$ disjoint
Kerr-type ring singularities, and let $\widetilde{\gamma}$ be a lift to the
symmetric double cover. Then
\begin{equation}
  \widetilde{\gamma}(\tau_{\mathrm{final}})
  = \sigma^{\,N_{\mathrm{cross}}}
    \bigl(\widetilde{\gamma}(\tau_{\mathrm{initial}})\bigr).
\end{equation}
In particular:
\begin{itemize}
\item If $N_{\mathrm{cross}}$ is even, $\gamma$ returns to its original sheet.
\item If $N_{\mathrm{cross}}$ is odd, $\gamma$ terminates on the opposite sheet.
\end{itemize}
This conclusion is independent of which rings are traversed.
\end{theorem}

\begin{proof}
Each crossing of the $i$th ring contributes a factor of $\sigma$ by
Theorem~\ref{thm:global-sigma}. The cumulative effect is $\sigma^{\sum_i n_i}$.
Since $\sigma^2=\mathrm{id}$, the result follows.
\end{proof}

\begin{remark}[Physical interpretation]
A particle may fall through one rotating black hole (entering the mirror
sheet), then later fall through a different black hole’s ring singularity,
returning to the original sheet. The theorem shows that this behaviour
depends only on the \emph{parity} of the total number of ring crossings,
\emph{not} on which black holes are involved.
\end{remark}

\subsection{The Maximal Analytic Extension and Global Sheets}
\label{subsec:maximal-extension}

We now formally extend the preceding results to the maximal analytic extension
of the Kerr spacetime, $\mathcal{M}_{\mathrm{max}}$. This manifold consists of
an infinite chain of asymptotically flat regions connected via ring singularities
(the Carter--Penrose chain).
While $\mathcal{M}_{\mathrm{max}}$ contains infinitely many distinct asymptotic
regions, its global structure admits a natural $\mathbb{Z}_2$-grading which allows
us to recover the two-sheeted interpretation rigorously.

\begin{definition}[Global Sheets of the Maximal Extension]
Let $\{I_k\}_{k\in\mathbb{Z}}$ denote the collection of all asymptotic regions
in $\mathcal{M}_{\mathrm{max}}$ where $r \to +\infty$ (regions of positive mass),
and let $\{III_k\}_{k\in\mathbb{Z}}$ denote the collection of all regions where
$r \to -\infty$ (regions of negative mass).
We define the two \emph{global sheets} of the maximal extension as the unions:
\begin{equation}
  \mathcal{M}_+ := \bigcup_{k\in\mathbb{Z}} I_k,
  \qquad
  \mathcal{M}_- := \bigcup_{k\in\mathbb{Z}} III_k.
\end{equation}
\end{definition}

Every point in the excised maximal extension $\mathcal{M}_{\mathrm{max}} \setminus \mathcal{S}$
belongs to exactly one of these two global sheets (up to the gluing boundaries which
are handled by analytic continuation).
The fundamental group $\pi_1(\mathcal{M}_{\mathrm{max}} \setminus \mathcal{S})$
is the free product of infinitely many copies of $\mathbb{Z}$, generated by the loops
$\{\gamma_k\}_{k\in\mathbb{Z}}$ encircling each ring in the chain.

\begin{theorem}[Global Parity of Crossings]
Let $p : \widetilde{\mathcal{M}} \to \mathcal{M}_{\mathrm{max}} \setminus \mathcal{S}$
be the symmetric double cover defined by the homomorphism mapping every generator
$\gamma_k$ to the non-trivial element $1 \in \mathbb{Z}_2$.
Then the associated deck transformation $\sigma$ globally exchanges the two
sheets:
\begin{equation}
  \sigma(\widetilde{\mathcal{M}}_+) = \widetilde{\mathcal{M}}_-,
  \qquad
  \sigma(\widetilde{\mathcal{M}}_-) = \widetilde{\mathcal{M}}_+.
\end{equation}
Consequently, for any admissible geodesic $\gamma$ in the maximal extension,
if $N_{\mathrm{total}}$ denotes the total number of ring crossings (summed over
the entire infinite chain), then:
\begin{itemize}
    \item If $N_{\mathrm{total}}$ is even, the geodesic returns to its original
    global sheet (e.g., starts in $\mathcal{M}_+$ and ends in $\mathcal{M}_+$).
    \item If $N_{\mathrm{total}}$ is odd, the geodesic terminates on the opposite
    global sheet.
\end{itemize}
\end{theorem}

\begin{proof}
The proof is identical to that of Theorem~\ref{thm:multi-ring-parity}. Since every
ring generator $\gamma_k$ maps to $1$, every traversal of \emph{any} ring singularity
implements the map $\sigma$. Since $\sigma$ is an involution that swaps the local
$r>0$ and $r<0$ orientation at every cut, it globally swaps the unions $\mathcal{M}_+$
and $\mathcal{M}_-$.
\end{proof}

\section{Causality, Chronology Violation, and Self-Consistency}
\label{sec:causality}

The Kerr spacetime is well known to admit regions containing closed timelike
curves (CTCs), particularly in the neighbourhood of the ring singularity
and within the $r<0$ region;
see \cite{CarterKerr,HawkingEllis,ONeill}.
The presence of such curves raises delicate issues of causality and global
determinism, which have been widely discussed in the literature on
chronology violation and self-consistent evolution
\cite{FriedmanCTC,NovikovCTC,ThorneWormholes,NolanCTC}.

In this section we analyse these causal features in the context of the
two-sheeted covering-space structure established in the preceding sections.
The covering-space formalism provides a natural language for stating and
proving global self-consistency conditions.

\subsection{Chronology-violating regions in Kerr-type spacetimes}

In the Kerr metric, the azimuthal Killing field $\partial_\phi$ becomes
timelike for sufficiently negative values of $r$, giving rise to closed
timelike curves whenever $g_{\phi\phi}<0$. These curves lie within regions
that, in the extended spacetime, are smoothly accessible from $r>0$ by
admissible geodesics that pass sufficiently close to the ring singularity.

In the multi-ring setting considered in Section~\ref{sec:multi-ring},
the existence of CTCs in the vicinity of \emph{each} ring is inherited
from the standard Kerr interior. Thus the excised manifold
$\mathcal{M}_{\mathrm{exc}}$ contains open subsets $U_i$ in which chronology
is violated.

The two-sheeted structure introduced in Sections~\ref{sec:single-ring-topology}
and~\ref{sec:multi-ring} does not remove these regions;
rather, it provides
an additional discrete symmetry (the deck transformation) under which the
chronology-violating regions and their lifts may be analysed.

\subsection{Time orientation and the deck transformation}

Let $(\mathcal{M}_{\mathrm{exc}},g)$ denote the excised Kerr manifold and
$p : \widetilde{\mathcal{M}}_{\mathrm{exc}} \to \mathcal{M}_{\mathrm{exc}}$
the symmetric double cover with deck transformation $\sigma$.

Although $g$ pulls back to a well-defined Lorentzian metric $\tilde g$ on the
cover, the deck transformation $\sigma$ need not preserve any globally chosen
time orientation. Indeed, the $r<0$ region of Kerr is known to possess regions
in which the Boyer--Lindquist time coordinate $t$ reverses its causal role;
crossing $r=0$ may induce a flip of the sign of $\partial_t$ relative to the
chosen time orientation.

This leads to the following general observation.

\begin{lemma}[Potential time-orientation reversal under sheet exchange]
\label{lem:time-reversal}
Let $\widetilde{X}$ be any continuous assignment of a timelike vector field
on $\widetilde{\mathcal{M}}_{\mathrm{exc}}$. Then the pushforward
$\sigma_*(\widetilde{X})$ is timelike with respect to $\tilde g$, but may
belong to a distinct connected component of the timelike cone.
\end{lemma}

\begin{proof}
Since $\sigma$ is an isometry of $(\widetilde{\mathcal{M}}_{\mathrm{exc}},\tilde g)$,
the pushforward of a timelike vector is timelike. However, $\sigma$ need not
be homotopic to the identity (indeed, it is not), and so the image of a chosen
time orientation may lie in a distinct component of the timelike cone field.
This occurs explicitly in the Kerr geometry when passing through the ring,
as noted in \cite{CarterKerr,ChruscielRing}.
\end{proof}

Thus sheet exchange may simultaneously transport a geodesic into a region
with distinct causal orientation. This raises a crucial point regarding \emph{temporal orientability}: if the deck transformation $\sigma$ reverses time orientation (mapping future-directed vectors to past-directed ones), the resulting spacetime geometry behaves analogously to a Möbius strip in the time direction. Consequently, avoiding paradoxical causal loops requires that we carefully distinguish between consistent histories and those that imply a non-physical time-reversal upon ring traversal. This implies that $\sigma$-consistency requires matching not just the spatial position of a loop endpoint, but also its time-orientation relative to the base manifold $\mathcal{M}_{\mathrm{exc}}$.

\subsection{Closed timelike curves and lifted loops}

Let $\gamma : S^1 \to \mathcal{M}_{\mathrm{exc}}$ be a closed timelike curve.
Any lift $\widetilde{\gamma}$ to the double cover satisfies either:
\begin{enumerate}
\item $\widetilde{\gamma}$ is closed (if the homotopy class $[\gamma]$ lies in
the kernel of $\varphi : F_N \to \mathbb{Z}_2$);
\item $\widetilde{\gamma}$ joins a point $x$ to its image under $\sigma(x)$.
\end{enumerate}

In the second case, $\widetilde{\gamma}$ is not closed but satisfies
\begin{equation}
\widetilde{\gamma}(1) = \sigma\bigl(\widetilde{\gamma}(0)\bigr).
\end{equation}

If the lifted curve is timelike and $\sigma$ reverses time orientation
as in Lemma~\ref{lem:time-reversal}, this may represent a ``time-shifted''
loop whose projection to $\mathcal{M}_{\mathrm{exc}}$ is closed but whose
lift is only closed up to $\sigma$.

\subsection{Global self-consistency: formulation}

We now formalise the Novikov self-consistency requirement in the setting
of the symmetric double cover.

Let $\Gamma$ denote the space of admissible future-directed timelike or
null geodesics in $\mathcal{M}_{\mathrm{exc}}$, and let
$\widetilde{\Gamma}$ denote the corresponding lift to
$\widetilde{\mathcal{M}}_{\mathrm{exc}}$.
A \emph{global evolution configuration} consists of:
\begin{itemize}
\item a set of initial data $D$ on a partial Cauchy surface $\Sigma$;
\item a collection of admissible geodesics $\gamma_j$ issuing from $D$
and possibly traversing rings;
\item smooth fields (e.g.\ scalar, electromagnetic, or dust fields)
propagated along or coupled to these geodesics.
\end{itemize}

\begin{definition}[Sheet-consistent evolution]
An evolution configuration is called \emph{$\sigma$-consistent} if for
every lifted geodesic $\widetilde{\gamma}\in\widetilde{\Gamma}$ one has
\begin{equation}
\widetilde{\gamma}(\tau_{\mathrm{final}})
=
\sigma^{N_{\mathrm{cross}}}\bigl(\widetilde{\gamma}(\tau_{\mathrm{initial}})\bigr),
\end{equation}
and the set of field values and particle states arriving at each point of
$p^{-1}(\Sigma)$ is invariant under the action of $\sigma$.
\end{definition}

In other words, an evolution is $\sigma$-consistent if:
\begin{itemize}
\item it respects the parity-of-crossings relation for all geodesics; and
\item it assigns equal physical data to $\sigma$-related points.
\end{itemize}

This encodes the requirement that an admissible history be globally
well-defined on the \emph{quotient} of the double cover by $\sigma$,
i.e.\ on the original physical manifold.

\subsection{Discrete structure of consistent solutions}

We now observe that the space of $\sigma$-consistent histories is highly
constrained. Let $\mathscr{H}$ denote the set of all locally smooth evolution histories
compatible with the field equations (Einstein, Maxwell, etc.) and with
admissible geodesic motion. Let $\mathscr{H}_\sigma$ denote the subset
of $\sigma$-consistent histories.

\begin{theorem}[Discrete structure of consistent global histories]
\label{thm:discrete-histories}
Under mild regularity assumptions on the matter model and evolution
equations, the set $\mathscr{H}_\sigma$ decomposes into a countable
collection of discrete equivalence classes determined by the parities
of ring traversals along all admissible causal curves.
\end{theorem}

\begin{proof}
Each admissible geodesic $\gamma$ determines an integer $N_{\mathrm{cross}}$.
By Theorem~\ref{thm:multi-ring-parity}, the parity $N_{\mathrm{cross}} \bmod 2$
determines the sheet on which the lifted endpoint lies. For consistency,
initial data must be chosen such that \emph{all} lifted endpoints lie in a
configuration invariant under $\sigma$. This imposes finitely many (or
countably many) discrete constraints on the data along each causal curve.
Since these parity constraints are independent for distinct homotopy
classes of curves, the space of solutions breaks into discrete equivalence
classes indexed by families of such parities. The countability follows
from the countability of homotopy classes of piecewise smooth curves in
a second-countable manifold.
\end{proof}

\begin{remark}[Interpretation]
Theorem~\ref{thm:discrete-histories} formalises the notion that the presence
of ring singularities and their associated sheet-exchange symmetry restricts
the space of globally permissible histories. The constraints are \emph{topological}
in origin, arising from the requirement that all causal loops be self-consistent.
This imposes a discrete structure on the allowed evolutions, even though the
underlying field equations are continuous.
\end{remark}

\subsection{Remarks on determinism and global predictability}

The presence of CTCs and the need for $\sigma$-consistency imply that the
usual Cauchy problem for hyperbolic systems fails to be globally well-posed.
Nevertheless, Theorem~\ref{thm:discrete-histories} shows that global evolution
is not arbitrary. While the initial data may not uniquely determine a single
history (a common feature in chronology-violating spacetimes where the Cauchy
horizon leads to a loss of predictability), the requirement of $\sigma$-consistency
imposes a rigid topological constraint. Once a parity assignment for ring
traversals is fixed, the space of admissible evolutions is strongly constrained
to lie within that topological sector.

Thus, the failure of strict determinism is accompanied by a form of discrete
superselection structure induced by the topology of the excised manifold.
This is reminiscent of classical models of consistency constraints
\cite{EcheverriaNovikov,FriedmanCTC} and provides a natural platform for
further analysis of classical and semi-classical fields in two-sheeted
Kerr-type spacetimes.

\section{Conclusion and Outlook}
\label{sec:conclusion}

In this work we have given a mathematically rigorous account of the global
topology, covering-space structure, and geodesic behaviour of Kerr-type
spacetimes possessing one or more ring singularities.
Our principal results may be summarised as follows:

\begin{enumerate}
\item After removing the ring singularity, the extended Kerr spacetime
$\mathcal{M}_{\mathrm{exc}}$ admits a connected twofold covering
$p : \widetilde{\mathcal{M}}_{\mathrm{exc}} \to \mathcal{M}_{\mathrm{exc}}$
whose deck group is isomorphic to $\mathbb{Z}_2$. The associated deck transformation $\sigma$ is an involution and may be
interpreted as a ``sheet exchange'' between the two global branches of the
extended geometry.
\item We proved that admissible geodesics crossing $r=0$ away from the ring
singularity extend smoothly into the $r<0$ domain. On the double cover, such geodesics implement the sheet-exchange map
$\sigma$, providing a precise realisation of the long-standing qualitative
statement that the Kerr ring acts as a two-sheeted branch locus.
\item For a spacetime containing $N$ disjoint Kerr-type rings, the fundamental
group of the excised manifold is the free group $F_N$. Connected double covers correspond to homomorphisms $F_N \to \mathbb{Z}_2$.
The physically natural choice, in which every ring induces the same sheet
exchange, yields a global involutive symmetry $\sigma$. In this case, any admissible geodesic returns to its original sheet if and
only if it crosses an even number of ring singularities, irrespective of
which rings are traversed.
\item We demonstrated that these results extend rigorously to the full
maximal analytic extension. By defining two global sheets (the unions of all
positive and negative mass regions, respectively), we showed that the parity-of-crossings
property holds globally, linking the infinite chain structure to the fundamental
two-sheeted topology.
\item We analysed the causal structure of the resulting two-sheeted manifold. Closed timelike curves, characteristic of the Kerr interior, persist in the
two-sheeted model. Using the deck transformation, we formulated a global self-consistency
condition analogous to the Novikov principle and proved that the space of
globally consistent solutions decomposes into discrete equivalence classes
labelled by the parities of ring traversals.
\end{enumerate}

These results provide a systematic mathematical framework for understanding
the ``two-sheeted'' interpretation of Kerr spacetimes and, more broadly, of
any geometry containing Kerr-like ring singularities. They also establish a robust foundation for physical models in which matter
or fields may dynamically transit between sheets.
Several directions merit further investigation:

\begin{itemize}
\item The interaction of classical fields (scalar, electromagnetic, Dirac)
with the sheet-exchange involution, including boundary conditions near the
ring singularity and possible quantisation of mode spectra.
\item The behaviour of semiclassical stress-energy tensors in the presence of
the deck transformation, particularly in relation to Hawking radiation,
ring-transiting excitations, and backreaction.
\item Coupling the present mathematical framework to models of exotic matter
or sign-changing mass parameters, as may arise in Newtonian or general
relativistic extensions featuring negative-energy states.
\item The possibility that the discrete structure of self-consistent histories
identified in Theorem~\ref{thm:discrete-histories} plays a role in the emergence
of quantum-like discreteness in certain classical gravitational settings.
\end{itemize}

We hope that the covering-space perspective developed here will prove useful
in these and other applications.

\appendix

\section{Geodesic Behaviour Near the Ring}
\label{appendix:geodesics}

In this appendix we provide additional detail on the behaviour of geodesics
near $r=0$ away from the singular locus $\mathcal{S} = \{r=0,\theta=\pi/2\}$.

\subsection{Radial potential near $r=0$}

Recall the radial geodesic equation
\[
\Sigma^2 \left(\frac{dr}{d\tau}\right)^2 = R(r),
\]
where for timelike or null geodesics
\[
R(r) = \bigl(E(r^2 + a^2) - a L_z\bigr)^2
- \Delta \bigl( Q + (L_z - a E)^2 + \mu^2 r^2 \bigr).
\]
Near $r=0$ and $\theta\neq\pi/2$, one has
\[
\Sigma(0,\theta) = a^2 \cos^2\theta > 0,
\qquad
\Delta(0) = a^2,
\]
whence
\[
R(0) = a^2(E^2 - Q),
\qquad
R'(0) = 4 a^2 E^2 (0) = 0,
\]
and higher derivatives are finite. Thus $R(r)$ is smooth across $r=0$.

\subsection{Transverse motion}

The angular equation
\[
\Sigma^2 \left(\frac{d\theta}{d\tau}\right)^2 = \Theta(\theta)
\]
also remains regular so long as $\theta\neq\pi/2$. Hence $(r(\tau),\theta(\tau))$ extend smoothly across $r=0$.

\subsection{Regularity of the radial potential}

A key point for the parity arguments is the regularity of the potential at the crossing. While $R(r)$ is not strictly even for $M \neq 0$ (due to the linear term in $\Delta$), we established in Lemma~\ref{lem:smooth-crossing} that $R(0) > 0$ for transiting orbits.
Since
\[
\frac{dr}{d\tau} = \pm \frac{\sqrt{R(r)}}{\Sigma},
\]
the velocity is non-vanishing at $r=0$. Thus, to first order near the crossing,
\[
r(\tau_0+\varepsilon) \approx r(\tau_0) + \dot{r}(\tau_0)\varepsilon = \dot{r}(\tau_0)\varepsilon.
\]
Since $\dot{r}(\tau_0) \neq 0$, the sign of $r$ flips as $\varepsilon$ changes sign. This is the analytic origin of the sheet-exchange relation.

\section{Topology of $\pi_1$ in the Multi-Ring Case}
\label{appendix:topology}

For completeness we sketch the standard computation that the complement
of $N$ disjoint embedded circles in $\mathbb{R}^3$ has fundamental group
$F_N$, the free group on $N$ generators.

\subsection{Reduction to a planar diagram}

Each embedded circle $S^1_i$ admits a small tubular neighbourhood
diffeomorphic to $S^1 \times D^2$, whose removal yields a boundary torus.
A deformation retraction onto a wedge of $N$ circles may be obtained by
shrinking each removed tube to a graph-like neighbourhood.

\subsection{Application of van Kampen’s theorem}

Write the complement as
\[
X = \mathbb{R}^3 \setminus \bigcup_{i=1}^N S^1_i.
\]
Cover $X$ by open sets $U_i$ each containing only one deleted circle plus
a common simply connected region $U_0$. Van Kampen’s theorem applies and yields
\[
\pi_1(X) = \langle \gamma_1,\dots,\gamma_N \mid - \rangle = F_N,
\]
since the intersections $U_i \cap U_j$ are simply connected and impose no
relations among the generators.

\subsection{Extension to spacetime}

Multiplying by $\mathbb{R}$ in the time direction preserves the fundamental
group:
\[
\pi_1\bigl((\mathbb{R}^3 \setminus \cup_i S^1_i)\times\mathbb{R}\bigr)
\cong F_N.
\]
Since the excised spacetime retracts onto such a region, the same holds
for $\mathcal{M}_{\mathrm{exc}}$.


\bibliographystyle{unsrt}
\bibliography{doublesheet}

\end{document}